\documentclass[article,nofootinbib]{revtex4-1}
\pdfoutput=1
\usepackage[utf8]{inputenc}
\usepackage{amsmath,amssymb}
\usepackage{graphicx}
\usepackage{subfig,xcolor}
\usepackage[font=small,labelfont=bf]{caption}
\usepackage[margin=1in]{geometry}
\usepackage{float}
\usepackage{amssymb}
\usepackage{bm}

\begin{document}

\title{Extreme scenarios: the tightest possible constraints on the power spectrum due to primordial black holes}
\author{Philippa S.~Cole}
\email{P.Cole@sussex.ac.uk}

\author{Christian T.~Byrnes}
\email{C.Byrnes@sussex.ac.uk}

\affiliation{Department of Physics and  Astronomy, University of Sussex, Brighton BN1 9QH, UK}	

\date{\today}

\begin{abstract}
Observational constraints on the abundance of primordial black holes (PBHs) constrain the allowed amplitude of the primordial power spectrum on both the smallest and the largest ranges of scales, covering over 20 decades from $1-10^{20}/ \rm{Mpc}$. Despite tight constraints on the allowed fraction of PBHs at their time of formation near horizon entry in the early Universe, the corresponding constraints on the primordial power spectrum are quite weak, typically ${\cal P}_\mathcal{R}\lesssim 10^{-2}$ assuming Gaussian perturbations. Motivated by recent claims that the evaporation of just one PBH would destabilise the Higgs vacuum and collapse the Universe, we calculate the constraints which follow from assuming there are zero PBHs within the observable Universe. Even if evaporating PBHs do not collapse the Universe, this scenario represents the ultimate limit of observational constraints. Constraints can be extended on to smaller scales right down to the horizon scale at the end of inflation, but where power spectrum constraints already exist they do not tighten significantly, even though the constraint on PBH abundance can decrease by up to 46 orders of magnitude. This shows that no future improvement in observational constraints can ever lead to a significant tightening in constraints on inflation (via the power spectrum amplitude). The power spectrum constraints are weak because an order unity perturbation is required in order to overcome pressure forces. We therefore consider an early matter dominated era, during which exponentially more PBHs form for the same initial conditions. We show this leads to far tighter constraints, which approach ${\cal P}_\mathcal{R}\lesssim10^{-9}$, albeit over a smaller range of scales and are very sensitive to when the early matter dominated era ends. Finally, we show that an extended early matter era is incompatible with the argument that an evaporating PBH would destroy the Universe, unless the power spectrum amplitude decreases by up to ten orders of magnitude.
\end{abstract}

%\begin{document}

\maketitle

\section{Introduction}

Primordial black holes (PBHs) can form from the collapse of large density fluctuations in the early Universe. If an overdensity of order unity in a given region reenters the Hubble sphere\footnote{`Hubble sphere' is used interchangeably with `horizon' throughout.} after inflation, then the region collapses to form a PBH with mass roughly equal to the mass within the Hubble sphere. 

The abundance of PBHs in our observable Universe today can constrain the primordial power spectrum, and hence models of inflation, on scales much smaller than are accessible via observations of the CMB and LSS (which provide the tightest constraints on the primordial power spectrum for scales between $k\sim10^{-3}-1\,{\rm Mpc^{-1}}$). Constraints on the abundance of PBHs are very tight at the time of their formation, due to their gravitational effects and the consequences of their evaporation if they were sufficiently light \cite{Carr:2009jm}. See \cite{Niikura:2017zjd} for some of the most up to date constraints on the abundance of PBHs\footnote{Presented in the context of whether PBHs can make up all of dark matter.}.  Despite constraints on PBHs being tight, the fact that PBH abundance and the power spectrum are related logarithmically during radiation domination %post-inflation 
means that even massively tightening the PBH abundance constraints does not translate into a great improvement on the constraints on the primordial power spectrum. Here we derive the tightest constraints possible on the primordial power spectrum given the most extreme constraints on the abundance of PBHs, i.e.~that there are none.  

Apart from providing the tightest possible future constraints on the primordial power spectrum\footnote{Assuming radiation domination and Gaussian initial perturbations.}, this extreme assumption is motivated by an argument that the decay of a PBH would destabilise the Higgs vacuum; hence PBHs of masses small enough that they would have decayed by today can't ever have formed \cite{Burda:2015isa,Burda:2016mou,Tetradis:2016vqb}. If they had they would have seeded the decay of the metastable Higgs vacuum, the Universe would have tunnelled to the true vacuum, and hence been destroyed. See \cite{Gorbunov:2017fhq,Canko:2017ebb} for limitations of this claim.

Since the logarithmic relation between PBH abundance and the power spectrum is the main cause of the suppression of the constraint, the most effective way of improving the constraint is by looking at  scenarios where the power spectrum amplitude is more sensitive to the PBH abundance. 
This is the case during an early matter dominated phase (see \cite{Kane:2015jia} and \cite{Acharya:2009zt} for motivations) where the relation between abundance and power spectrum is instead given by a power law \cite{Polnarev1981,Harada:2016mhb}. This means that for an observed abundance of a particular PBH mass, the constraint on the primordial power spectrum is tightened by many orders of magnitude, suggesting that the best constraint possible will come from such a scenario where the equation of state is at its minimum non-negative value, $\omega=0$. Models with $-\frac{1}{3}<\omega<0$ are rarely considered, and $\omega<-1/3$ corresponds to inflation. In terms of the worst constraints, according to \cite{Harada:2013epa}, the value of $\omega$ for which it is most difficult to produce PBHs is $\omega=1/3$ due to the value of critical overdensity being at its maximum. The threshold density then decreases as $\omega$ approaches 1. However, this is different to the results of \cite{Musco:2012au} whose value of critical overdensity increases with $\omega$ for $0<\omega<0.6$ (other values of $\omega$ were not simulated) suggesting that the larger $\omega$ is, the more difficult it is to produce PBHs. 

Generally, calculations done in the radiation-dominated scenario assume spherical symmetry of the collapsing region (which makes collapse as likely as possible) but those done in the matter-dominated scenario do not \cite{Kuhnel:2016exn}. Full numerical simulations of the collapse of density fluctuations to form PBHs are required to gain a complete understanding of the process. The critical density required for a region to collapse and form a PBH also depends on the density profile. Additionally, phenonema such as non-Gaussianity could have  a significant influence on the abundance of PBHs since they are formed from rare, large overdensities. Such large fluctuations are susceptible to changes in the tail of the fluctuation distribution caused by the amount of non-Gaussianity present \cite{Young:2013oia}. 
 We will not show explicitly the effects of non-Gaussianity on our results but its potential effect should be kept in mind. Furthermore for simplicity, we will assume that all PBHs form with the same mass for a given time. In reality, PBH masses depend on the PBH mass function (see \cite{Carr:2017jsz}  for a recent update) which would affect the scale at which primordial power spectrum constraints are correlated to. 

The paper is laid out as follows: in section \ref{Ian Moss} we will discuss the argument in \cite{Burda:2015isa,Burda:2016mou} for zero PBHs due to the Higgs instability and calculate constraints on the primordial power spectrum in this case. In section \ref{Matter} we will look at motivations for a matter dominated phase and find constraints on the primordial power spectrum for different durations of matter domination prior to BBN. Finally, in section \ref{section:mossmatter} we will combine these two frameworks and see the result of their co-existence which is shown to be in agreement with the results of \cite{Gorbunov:2017fhq}.

\section{No PBH formation during radiation domination}\label{Ian Moss}

It is believed that the electroweak vacuum is metastable, depending on the mass of the top quark, with a lifetime longer than the present age of the Universe \cite{Anderson:1990aa,Arnold:1991cv}. The notion that impurities initiate phase transitions gives rise to the idea that natural inhomogeneities such as PBHs may be capable of seeding rapid vacuum decay from the metastable vacuum to the true vacuum. If this had happened before today, the process would have had catastrophic consequences for the Universe. This is proposed and explored in \cite{Burda:2015isa,Burda:2016mou,Gregory:2013hja,Burda:2015yfa}. See \cite{Gorbunov:2017fhq,Canko:2017ebb} for limitations of this claim.

The conclusion drawn from this argument is that since the Universe has not been destroyed,  no PBHs with masses small enough such that they would have already decayed can ever have formed. This is one motivation for no light PBHs forming, however our results will also represent the limit of improving observational constraints on the abundance of PBHs to their absolute tightest. For a radiation dominated background, assuming that the PBH mass, $M_{\rm PBH}$, is of the same order as the horizon mass, the relationship between PBH mass and horizon entry time, $t_{i}$, which is approximately the formation time during radiation domination, is given by
\begin{equation}\label{hormass}
    M_{\rm PBH}=\gamma10^{15}\left(\frac{t_{i}}{10^{-23}{\rm s}}\right){\rm g},
\end{equation}
where $\gamma$ is the ratio between horizon mass and PBH mass  \cite{Carr:2009jm}. The mass of a PBH which would just be decaying today is around $10^{15}$g \cite{Carr:2009jm} so from equation (\ref{hormass}) we can say that PBHs which could have formed at or before $\sim10^{-23}$ seconds would have catalysed the rapid vacuum decay, and therefore never existed. We will use this bound on abundance of PBHs being zero to find the tightest possible constraint on the primordial power spectrum originating from the non-detection of PBHs.

\subsection{Constraint relations}
The abundance of PBHs is usually described by the PBH mass fraction:
\begin{equation}\label{beta}
\beta(M_{\rm PBH})=\frac{\rho_{\rm PBH}(M_{\rm PBH})}{\rho_{\rm tot}},
\end{equation}
which denotes what fraction of the total energy density of the Universe is contained in regions dense enough to form PBHs at horizon entry, where $\rho_{\rm PBH}$ is the energy density contained within PBHs, and $\rho_{\rm tot}$ is the total energy density of the Universe. During radiation domination, PBHs form shortly after horizon entry. As will be seen later, during matter domination PBH formation occurs a significant time after horizon entry.

Assuming radiation domination, in order for a region of space-time to collapse and form a PBH, the smoothed density contrast at horizon crossing, $\delta(R)$, needs to exceed some critical level of over-density, $\delta_c$, which is of order 1. If the initial density perturbations have a Gaussian distribution then the pdf of the smoothed density contrast is given by: 
\begin{equation}
P(\delta(R))=\frac{1}{\sqrt{2\pi}\sigma(R)}{\rm exp}\left(\frac{-\delta^2(R)}{2\sigma^2(R)}\right)
\end{equation}
where $\sigma(R)$ is the mass variance:
\begin{equation}
\sigma^2(R)=\int^\infty_0\widetilde{W}^2(kR)\mathcal{P}_\delta(k)\frac{dk}{k}.
\end{equation}
$\mathcal{P}_\delta(k)$ is the primordial power spectrum of $\delta$ at horizon entry, describing how the overdensities and underdensities are distributed according to scale, $\delta=(\rho-\bar{\rho})/\bar{\rho}$ is the comoving density contrast and $\widetilde{W}$ is a smoothing function \cite{Byrnes:2012yx}. 

The PBH mass fraction in equation (\ref{beta}) is related to the probability density function as 

\begin{eqnarray}
\beta(M_{\rm PBH})
&=&\frac{2}{\sqrt{2\pi}\sigma(R)}\int^\infty_{\delta_c}\exp\left(\frac{-\delta^2(R)}{2\sigma^2(R)}\right)d(\delta(R))\nonumber\\
&=&{\rm Erfc}\left(\frac{\delta_{\rm c}}{\sqrt{2}\sigma(R)}\right),
\end{eqnarray}
where Erfc is the complementary error function and we have included the Press-Schechter\footnote{Since the factor of 2 ends up in the argument of the complementary error function in the expression for the power spectrum, including it or not only changes the results for the power spectrum by 1-2\%.} factor of 2.
By inverting this expression, it is possible to find constraints on the mass variance given constraints on $\beta$. We can hence find constraints on the power spectrum via $\sigma^2\sim\cal{P}$ if one assumes that the power spectrum is independent of $k$. We will therefore be using this expression:

\begin{equation}\label{Erfc}
\mathcal{P}_{\delta}\sim\sigma^2=\left(\frac{\delta_c}{\sqrt{2}{\rm InverseErfc}(\beta)}\right)^2
\end{equation}
to plot our constraints on the power spectrum\footnote{In this paper we always plot the constraints assuming a monochromatic mass spectrum of PBHs.}.

In order to construct a constraint on $\beta$ that represents there being zero PBHs for a certain range of masses, we can model the observable Universe as a cube of volume $L^3$, made up of $N_l$ smaller cubes each with volume $l^3$. These small cubes represent the size of the patches that may have collapsed to form black holes in the early Universe. In our observable Universe today, if there is just one black hole that formed in the early Universe, then a patch of size $l$ will have been overdense enough to collapse at the time that the patch reentered the horizon. 

If less than one out of all of these patches (i.e.~none of them) contain a black hole then
\begin{equation}\label{betaconstr}
\beta < \frac{1}{N_l}=\left(\frac{l}{L}\right)^3.
\end{equation}
Our constraint on the primordial power spectum therefore becomes: 
\begin{equation}
\mathcal{P}_{\delta}<\left(\frac{\delta_c}{\sqrt{2}{\rm InverseErfc}(\frac{1}{N_l})}\right)^2.
\end{equation}

\subsection{Relevant scales for Higgs stability argument}\label{section:scales}
As we saw from equation (\ref{hormass}), a PBH with mass $10^{15}{\rm g}$ formed at $10^{-23}$s is the largest and latest PBH that could have seeded rapid vacuum decay. We want to work out the physical size of the overdense region that would have needed to collapse to form a black hole of such mass, how large that region has expanded to today, and how many patches of that size there are in the Universe today. This will provide the threshold scale for the range of PBH masses capable of seeding rapid vacuum decay.

Assuming radiation domination, the physical size (length), $l_{\rm phys}$, of the horizon at horizon entry time is given by 

\begin{equation}\label{physhorizon}
l_{\rm phys}=\int^{t_{\rm i}}_0\frac{a_{i}}{a}dt=\int^{t_{i}}_0\left(\frac{t_{i}}{t}\right)^{\frac{1}{2}}dt=2t_{i}
\end{equation}
setting the speed of light to $c=1$. 
The scale that this has grown to today, i.e.~the comoving scale, can be found by multiplying the physical scale by the ratio between the scale factor today (defined as $a_0=1$) and the scale factor at the time that the black hole formed, i.e.~

\begin{equation}\label{scaletime}
\frac{a_0}{a_{i}}=\frac{a_{\rm eq}}{a_{i}}\frac{a_0}{a_{\rm eq}}\approx\left(\frac{t_{\rm eq}}{t_{i}}\right)^{\frac{1}{2}}\left(\frac{t_0}{t_{\rm eq}}\right)^{\frac{2}{3}},
\end{equation}
where $a_{\rm eq}$ is the scale factor at matter-radiation equality\footnote{We do not account for any late time dark energy domination in the evolution of scales nor $\beta$ throughout our work. The duration of this era is too short to have a dominant effect over some of the other uncertainties in the calculations.}. Equation (\ref{scaletime}) is approximate since we have assumed that the transition between radiation domination and matter domination is instantaneous for simplicity, and we have therefore also neglected the lesser effect of the change in degrees of freedom, $g_*$, between $t_i$ and $t_{\rm eq}$. Its effect would contribute a factor of $g_*^{-\frac{1}{12}}$ \cite{Liddlebook} to $a_0/a_i$, which would be at most order 1, since it decreases from order 100 to $\sim3$ between $t_i$ and $t_{\rm eq}$. We therefore have that the size of the horizon at a given horizon entry time has grown to a size today, $l_i\rvert_{\rm t_0}$, and comoving scale, $k_i$, given by: 
\begin{align}
    l_i\rvert_{\rm t_0}=2t_i{c}\left(\frac{t_{\rm eq}}{t_{i}}\right)^{\frac{1}{2}}\left(\frac{t_0}{t_{\rm eq}}\right)^{\frac{2}{3}}=\frac{2\pi}{k_i}
\end{align}
Inserting values for radiation-matter equality and the age of the Universe ($t_{\rm eq}\simeq10^{12}{\rm s}$,
$t_0\simeq4\times10^{17}{\rm s}$), and using $t_{i}\sim10^{-23}$s, we find that the physical size converted to Mpc of the region today is $l_i\rvert_{\rm t_0}\sim10^{-15}$Mpc, and the corresponding scale is $k_i\sim10^{16}$Mpc$^{-1}$. This represents the largest horizon scale of a PBH which would have decayed by today.

In order to determine the smallest scale that we can probe with PBHs in this scenario, we need to find the scale that left the horizon just before inflation ended and reentered immediately afterwards. If $H$ can be approximated as being constant during inflation, which is typically the case for small-field inflation, then the number of e-folds that occur between the time that today's horizon scale left the Hubble sphere during inflation and the end of inflation, $\Delta N$ is 
\begin{align}
    \Delta{N}=\ln\left(\frac{k_{\rm end}}{k_0}\right) \qquad
    \Rightarrow \;\;\; k_{\rm end}=k_0e^{\Delta{N}}.
\end{align}
Taking a value of $\Delta{N}$ within the expected range \cite{Liddle:2003as}, for example 60, and $k_0=a_0H_0\simeq2.3\times10^{-4}\,{\rm Mpc^{-1}}$  with $H_0=68\,{\rm km\,s^{-1}\,Mpc^{-1}}$ \cite{Ade:2015xua}, the smallest scale to leave the horizon just before the end of inflation is $k_{\rm end}\simeq2.6\times10^{22}\,{\rm Mpc^{-1}}$. 

If $H$ cannot be approximated as being constant during inflation, which is the case for large-field models of inflation, then the above expression becomes less accurate \cite{Liddle:2003as}. For example in quadratic inflation, with $V=m^2\phi^2/2$, the smallest scale is instead given by
\begin{equation}
    k_{\rm end}=\frac{\sqrt{2}k_0}{\sqrt{2\Delta{N}+1}}e^{\Delta{N}}.
\end{equation}
For $\Delta{N}=60$, this decreases $k_{\rm end}$ by about a factor of $10$.

\subsection{Results}
\begin{figure}
\centering
\includegraphics[width=15cm]{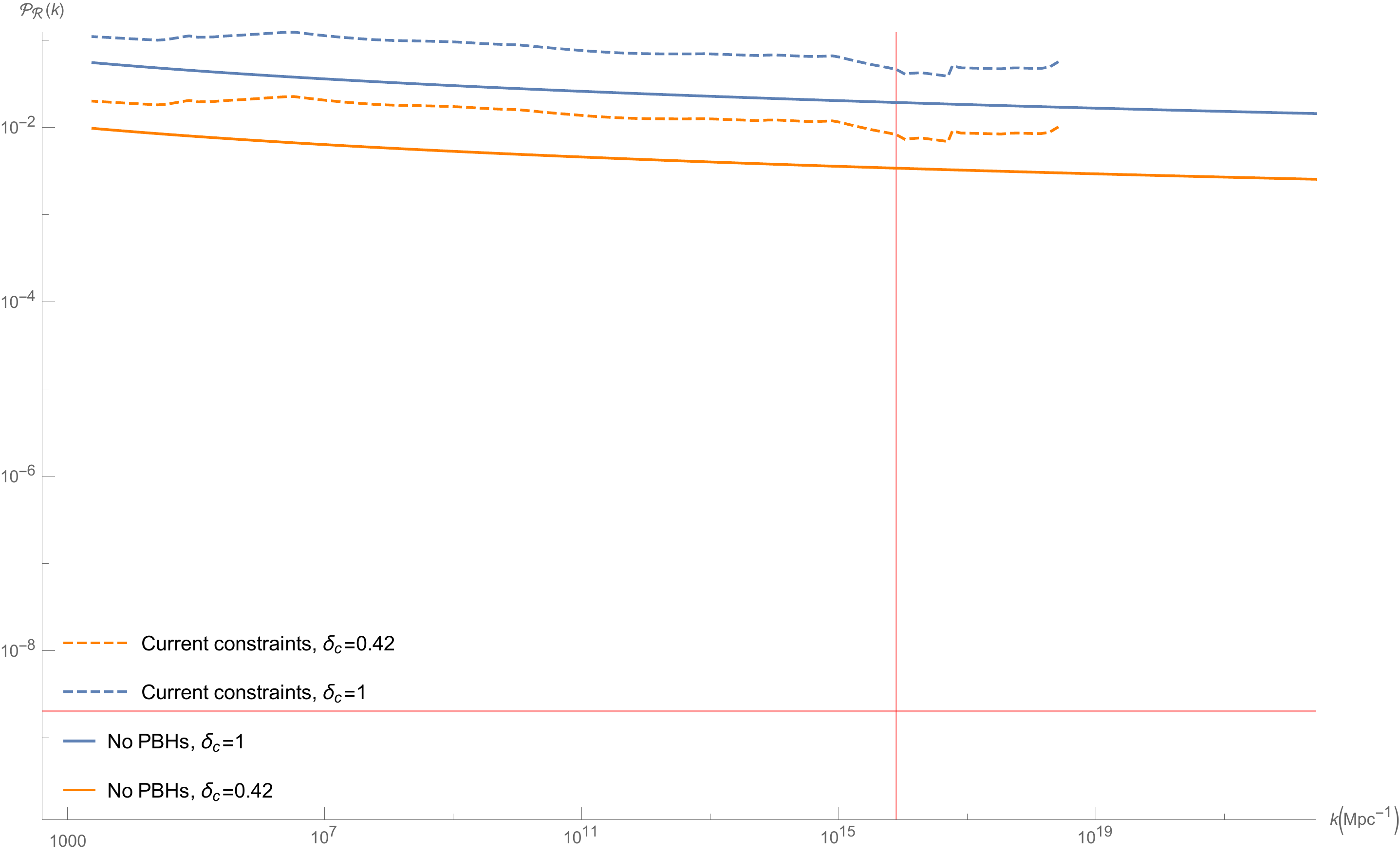}\captionof{figure}{Power spectrum constraints from PBH formation in radiation domination. The dashed lines represent the current constraints with two different values of critical level of overdensity. The solid lines represent the constraints if no PBHs form. The red vertical line represents the scale of a PBH that would just have decayed by today, so everything to the right of this line is the constraint due to the argument in \cite{Burda:2015isa,Burda:2016mou}. The horizontal red line is drawn at $\mathcal{P}_\mathcal{R}(k)=2\times10^{-9}$. The smallest value of $k$ plotted corresponds to $M_{\rm PBH}=10^{40}{\rm g}$, whilst the largest value of $k$ plotted corresponds to the smallest scale to reenter the horizon post-inflation as found in section \ref{section:scales} with $\Delta{N}=60$. We used $\gamma=0.2$ to plot these results \cite{Inomata:2017okj,Carr:1975qj}. 
}\label{fig:ianmoss}
\end{figure}

We can now plot the consequences for the primordial power spectrum of no PBHs forming on all scales and compare this to the current constraints. For scales smaller than $k\sim10^{16}{\rm Mpc^{-1}}$, indicated by everything to the right of the vertical red line in figure \ref{fig:ianmoss}, the constraint plotted is a consequence of the claim in \cite{Burda:2015isa,Burda:2016mou}. For scales larger than $k\sim10^{16}{\rm Mpc^{-1}}$, we do not have any evidence to suggest that no PBHs can have formed, but the plot demonstrates the effect on the constraints of the power spectrum if this were to be the case. It therefore also provides the tightest possible future constraint from PBHs, assuming there are none.

Using (\ref{betaconstr}) and $k_0=2\pi/L$, $\beta$ as a function of scale is
\begin{equation}
    \beta\simeq1.2\times10^{-11}(k\,{\rm Mpc})^{-3}.
\end{equation}
Then with equation (\ref{Erfc}) we calculate the constraint on the power spectrum $\mathcal{P}_\delta(k)$ against scale $k$, measured in Mpc$^{-1}$. Converting from the comoving density contrast $\delta$ to the comoving curvature perturbation $\mathcal{R}$ with 
\begin{equation}
    \delta=\frac{2(1+\omega)}{5+3\omega}\left(\frac{k}{aH}\right)^2\mathcal{R}
\end{equation}
evaluated at horizon crossing so $k=aH$ \cite{Green:2004wb} implies 
\begin{equation}\label{eqn:deltaR}
    \mathcal{P}_\mathcal{R}=\left(\frac{5+3\omega}{2(1+\omega)}\right)^2\mathcal{P}_\delta.
\end{equation}
During radiation domination ($\omega=1/3$), $\mathcal{P}_\mathcal{R}=(81/16)\mathcal{P}_\delta$. The resulting plot of $\mathcal{P}_\mathcal{R}$ against $k$ is shown in figure \ref{fig:ianmoss}. Plotting $\mathcal{P}_\mathcal{R}$ instead of $\mathcal{P}_\delta$ allows a comparison to be made with Planck's observed value of the amplitude of the primordial power spectrum on large scales, $A_s=2\times10^{-9}$ \cite{Ade:2015xua}.
Figure \ref{fig:ianmoss} shows that despite the constraints on $\beta$ being as extreme as they can be, the constraint on the primordial power spectrum only improves by around half an order of magnitude in comparison to the current constraints given in \cite{Carr:2009jm}. The value of $\delta_c$ chosen is important since the constraint varies with the square of this value, and this has much more of an effect than any variation in $\beta$. However, the effect of $\delta_c$ is the same for both the current constraints and our new constraints, so the improvement from the current constraints to the constraints based on there being no PBHs is the same for whichever value of $\delta_c$ is chosen. For our choice of $\delta_c=0.42$ \cite{Harada:2015yda} (see also \cite{Musco:2004ak,Shibata:1999zs} which derive a similar value of $\delta_c$), the tightest constraint $\mathcal{P}_\mathcal{R}\simeq2.5\times10^{-3}$ is reached at the scale $k\simeq2.6\times10^{22}\,{\rm Mpc^{-1}}$, which is the smallest scale to reenter the horizon post-inflation (found in section \ref{section:scales} with $\Delta{N}=60$) and the largest value of $k$ plotted in figure \ref{fig:ianmoss}. This shows that we cannot do any better than $\mathcal{P}_\mathcal{R}\simeq2.5\times10^{-3}$ from only knowing the constraint on $\beta$ and taking the smallest scale to reenter the horizon post-inflation to be the value found in section \ref{section:scales} with $\Delta{N}=60$. These calculations assume spherical collapse, and as pointed out by \cite{Akrami:2016vrq}, constraints on the power spectrum from non-detection of PBHs are very uncertain due to effects such as non-spherical collapse and critical collapse. Finally, we note that because PBHs form deep into the tail of the probability distribution, the effect of non-Gaussianity can have a much larger effect on the constraints than changing the constraints on $\beta$ from it being it's largest possible value such that PBHs form all of dark matter right down to assuming there are no PBHs, potentially changing the constraints by two orders of magnitude \cite{Lyth:2012yp,Byrnes:2012yx,Shandera:2012ke,Bugaev:2013vba,Young:2014oea,Young:2015cyn}. 

\section{Early matter-dominated phase}\label{Matter}
\subsection{Motivations}
In order to achieve tighter constraints on the primordial power spectrum  using PBH abundance constraints, we require a scenario where the power spectrum depends on $\beta$ more sensitively than logarithmically. This is the case during an early matter-dominated phase which can be caused by a scalar field which dominates the background energy density (e.g.~the inflaton or curvaton) oscillating in a quadratic potential \cite{Kane:2015jia,Acharya:2009zt}. PBH formation in an early matter-dominated phase has been studied in various previous works \cite{Harada:2016mhb,Polnarev1981,Polnarev:1986bi,Georg:2016yxa,Georg:2017mqk,Gorbunov:2017fhq,Carr:2017edp}, where it was shown that the relationship is governed by a power law instead of a logarithmic function. The exponentially enhanced probability of formation is due to the fact that the Jean's pressure which would normally halt PBHs from forming on sub-Hubble scales during radiation domination vanishes in matter domination, and so PBHs are able to form more easily.

\subsection{PBH formation likelihood and power spectrum constraints}
Based on the results of \cite{Harada:2016mhb}, the expression relating PBH abundance and the mass variance $\sigma$ (and hence the power spectrum via $\mathcal{P}_\delta\sim\sigma^2$) is

\begin{equation}\label{eqn:Harada}
\beta_0\simeq0.056\sigma^5,
\end{equation}
where $\beta_0$ is the PBH abundance fraction defined at the time of formation. This expression does not assume spherical symmetry in the initial density profile. Using data from the plot of $\rho_{\rm PBH}/\rho_{\rm DM}$ against $M_{\rm PBH}$ from \cite{Carr:2017edp} (which is in turn collated from constraints due to evaporation \cite{Carr:2009jm}, femto-lensing of gamma-ray bursts \cite{Barnacka:2012bm}, neutron-star capture \cite{Capela:2013yf}, white dwarf explosions \cite{Graham:2015apa}, microlensing \cite{Niikura:2017zjd,Tisserand:2006zx,Allsman:2000kg}, Planck results \cite{Ali-Haimoud:2016mbv}, survival of stars in Segue I \cite{Koushiappas:2017chw} and Eridanus II \cite{Brandt:2016aco}, and distribution of wide binaries \cite{Monroy-Rodriguez:2014ula}),  it is possible to scale the observed constraints on PBH abundance such that they include a period of evolution in the matter dominated phase. Taking constraints on $\rho_{\rm PBH}/\rho_{\rm DM}$ from \cite{Carr:2017edp}, we can relate them to constraints on $\beta(M)$ via
\begin{equation}\label{eqn:Tommi}
\beta(M)=\frac{\rho_{\rm PBH}}{\rho_{\rm DM}}\Omega_{\rm DM}\left(\frac{M_i}{M_\odot}\right)^\frac{1}{2}\left(\frac{M_\odot}{M_{\rm eq}}\right)^\frac{1}{2}
\end{equation}
where $\Omega_{\rm DM}\simeq0.26$ \cite{Ade:2015xua} and the horizon mass at matter-radiation equality will be taken as $7\times10^{50}{\rm g}$ \cite{Green:2004wb}.

\begin{figure}
\centering
\includegraphics[width=10cm]{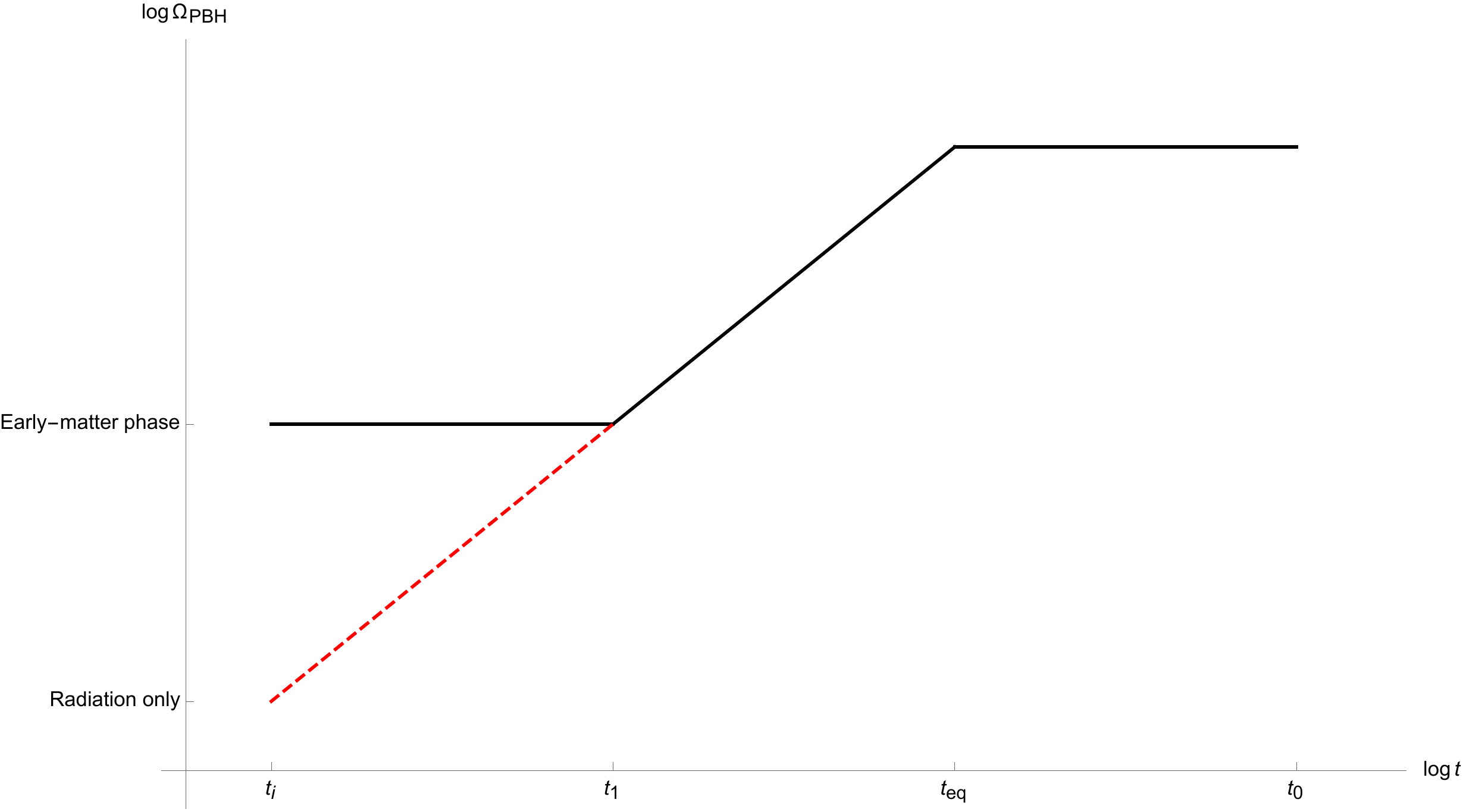}\captionof{figure}{Sketch of how $\Omega_{\rm PBH}={\rho_{\rm PBH}}/{\rho_{\rm tot}}$ scales with time depending on the periods of matter and radiation domination, where $\beta$ is equal to $\Omega_{\rm PBH}$ at horizon entry. The solid black line shows how $\Omega_{\rm PBH}$ evolves given an early matter dominated phase that lasts until $t_1$, followed by a radiation dominated phase where $\Omega_{\rm PBH}$ grows as $a$ and then the later matter dominated phase. The red dashed line shows where $\beta$ from \cite{Carr:2009jm} is evaluated with no early matter dominated phase. }\label{fig:betascaling}
\end{figure}

The values of $\beta$ found via equation (\ref{eqn:Tommi}) are calculated under the assumption that PBHs were forming in a radiation dominated phase. During radiation domination $\rho_{\rm PBH}/\rho_{\rm tot}$ grows like the scale factor ${a}$, whereas during matter domination ${\rho_{\rm PBH}}/{\rho_{\rm tot}}$ stays constant. In order to account for the period of time between formation and the end of the early matter dominated phase where ${\rho_{\rm PBH}}/{\rho_{\rm tot}}$ will remain constant instead of growing as ${a}$ we need to scale the observed value for $\beta$ by $a$. Figure \ref{fig:betascaling} gives a graphical depiction of this difference in scaling.  The relation between $\beta(M)$ from equation (\ref{eqn:Tommi}) and $\beta_0$ from equation (\ref{eqn:Harada}) is then
\begin{align}\label{eqn:beta0}
\beta_0=\frac{a_1}{a_{i}}\beta(M) \approx\left(\frac{t_1}{t_{i}}\right)^{\frac{2}{3}}\beta(M),
\end{align}
where the subscript `1' refers to the end of the early matter-dominated phase (and hence the beginning of the radiation dominated phase), and the subscript `$i$' refers to the time that the overdensity enters the Hubble sphere. As for equation (\ref{scaletime}), we have neglected the effect of the degrees of freedom in (\ref{eqn:beta0}). The change in degrees of freedom will however be even smaller here since it changes at most from order 100 to order 10 between $t_i$ and $t_1$, so the factor of $g_*^{-\frac{1}{12}}$ is negligible. In order to relate the PBH masses to different scales, we will take $\gamma=1$ since the precise value is uncertain for matter domination and assume that the mass of the resultant black hole is of the order of the mass of the horizon as the overdensity enters: 

\begin{align}
M_{\rm PBH}\sim{M_{\rm hor}}=\frac{c^3}{2GH(t_{i})}=\frac{3{c^3}t_{i}}{4G},
\end{align}
where we have used $H(t_{i})={2}/({3t_{i}})$ during matter domination.
The physical radius of the horizon at the time of horizon crossing is given by the gravitational radius, 
\begin{equation}
r_{\rm phys}=\frac{2GM_{\rm hor}}{c^2}=\frac{2GM_{\rm PBH}}{{c^2}},
\end{equation}
which has expanded to a radius today given by
\begin{equation}
r_i\rvert_{t_0}=r_{\rm phys}\left(\frac{a_0}{a_{i}}\right)={r}_{\rm phys}\left(\frac{t_1}{t_{i}}\right)^{\frac{2}{3}}\left(\frac{t_{\rm eq}}{t_{1}}\right)^{\frac{1}{2}}\left(\frac{t_0}{t_{\rm eq}}\right)^{\frac{2}{3}},
\end{equation}
where $t_1$ is the time that the early matter-dominated phase ends.
The scale today of PBHs with a particular mass at formation is:
\begin{equation}\label{eqn:know}
k_i(M_{\rm PBH})=\frac{2\pi}{r_i\rvert_{t_0}}=\frac{{2\pi c^2}}{2GM_{\rm PBH}}\left(\frac{t_1}{t_{i}}\right)^{-{\frac{2}{3}}}\left(\frac{t_{\rm eq}}{t_{1}}\right)^{-{\frac{1}{2}}}\left(\frac{t_0}{t_{\rm eq}}\right)^{-{\frac{2}{3}}}.
\end{equation}
Comparing this to the scale we found for a PBH that would have just decayed by today with mass $10^{15}$g, we find that 
\begin{equation}
k_i(10^{15}{\rm g})\simeq\frac{\rm (1\,s)^\frac{1}{6}}{{t_1^{\frac{1}{6}}}}\times5\times10^{12}\,{\rm Mpc^{-1}}.
\end{equation}
Taking the longest possible duration for the matter phase so that it lasts right up until BBN at $t_1=1$ second, we find $k_i(10^{15}{\rm g})\simeq5\times10^{12}{\rm Mpc^{-1}}$ which is around 4 orders of magnitude larger than if there was no early matter era. Choosing $t_1=10^{-23}$ seconds instead so that the evolution of the PBH is solely within the radiation dominated phase, we see that $t_1=t_{i}$ and $k_i\sim10^{16}{\rm Mpc^{-1}}$ which matches the value found in section \ref{Ian Moss}. This consistency check assumes that the matter phase ends and transitions to a radiation phase instantaneously, with the PBH forming as this happens - we will discuss the validity of this assumption as well as the collapse time of the PBHs shortly.

We will plot power spectrum constraint against  lengthscale using the inverse of equation (\ref{eqn:Harada}):
\begin{equation}\label{eqn:powermatter}
\mathcal{P}_\delta=\left(\frac{\beta_0}{0.056}\right)^{\frac{2}{5}},
\end{equation}
and (\ref{eqn:know}). Both of these quantities depend on the PBH mass so we take data from the plot in \cite{Carr:2017edp} at PBH masses from $10^{10}{\rm g}$ to $10^{40}{\rm g}$. Each mass has a corresponding value for $\beta$ which we find from equation (\ref{eqn:Tommi}), and then scale to $\beta_0$ via equation (\ref{eqn:beta0}). This is the value that is substituted into equation (\ref{eqn:powermatter}) to give values of the power spectrum corresponding to the scale for each PBH mass from equation (\ref{eqn:know}).

Since the power spectrum and the scale both depend on when the early matter dominated era ends, each chosen value of $t_1$ will result in a different constraint on the power spectrum. Additionally, each value of $t_1$ will determine the largest and smallest scales with observable consequences for PBH masses $10^{10}-10^{40}{\rm g}$ that can enter the horizon during the matter-dominated phase. We always assume that the early matter era begins before the horizon mass has grown to $10^{10}{\rm g}$. There will need to be enough time between horizon entry and turnaround time for initial overdensities $\delta_i$ to grow to order 1 (or the chosen $\delta_c$) if they are to collapse to form PBHs, so the constraints will weaken on larger scales that enter the horizon close to the end of the matter dominated phase as there won't be enough time for small initial density perturbations to grow before the matter-dominated phase ends.

Looking first at the largest and smallest scales that can enter the horizon before the end of the matter-dominated phase, we see that the smallest mass we use from \cite{Carr:2017edp} is $10^{10}\,{\rm g}$, which corresponds to a horizon entry time of $t_{i}\simeq3\times10^{-29}\,{\rm seconds}$. This means that the earliest time $t_1$ to the nearest power of 10 that matter domination can end and there still have been time for PBHs within the mass ranges we have data for to form  is $t_1=10^{-28}\,{\rm seconds}$. 

The smallest scale that enters the horizon post-inflation will be given by $k_i(M_{\rm PBH})$ evaluated at $M_{\rm PBH}=10^{10}\,{\rm g}$ for each value of $t_1$. The largest scale to enter the horizon before the phase transition from matter domination to radiation domination is determined by the value of $k_i(M_1)$ with $M_{1}$ given by
\begin{equation}
M_{1}=\frac{3{c^3}t_{1}}{4G}.
\end{equation}

\begin{figure} 
\centering
\includegraphics[width=15cm]{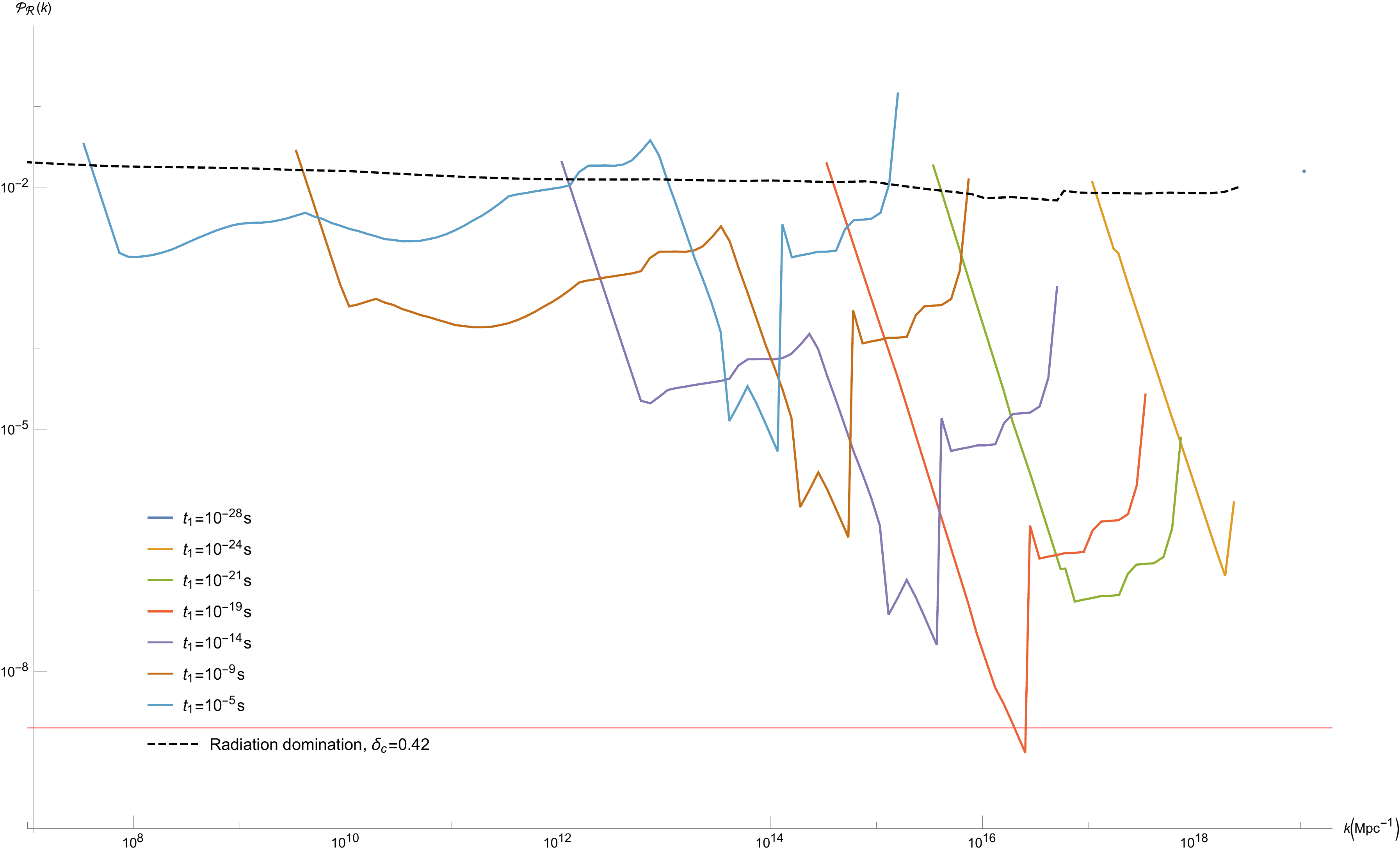}\captionof{figure}{Power spectrum constraints from PBH formation in matter domination for values of $t_1$ from $10^{-28}\,{\rm seconds
}$, represented by the single rightmost point, to $10^{-5}\,{\rm s}$, represented by the pale blue line that reaches the largest scales. The horizontal red line is drawn at $\mathcal{P}_\mathcal{R}(k)=2\times10^{-9}$. The dashed black line shows the constraints from PBHs which formed during radiation domination. }\label{fig:gorbonovmatter}
\end{figure}

If the overdensities were to collapse instantaneously after horizon entry then $k_i(10^{10}\,{\rm g})$ and $k_i(M_1)$ would determine the range of scales to be plotted for each value of $t_1$. However, overdensities do not collapse instantaneously to form PBHs after crossing the horizon. Instead, there needs to be enough time between horizon crossing and the end of the matter dominated phase for the overdensity to grow to order 1 and begin to collapse at the time of the `turnaround' \cite{Gorbunov:2017fhq}. We impose for simplicity the requirement that the overdensity must reach turnaround before the end of the early matter dominated phase if it is to collapse, so the scale factor will need to grow by a factor of $\delta_i^{-1}$ between $t_i$ and $t_{1}$ for the density fluctuation entering the horizon at $t_i$ to have had time to grow to order 1:
\begin{align}
    \delta(M_i)\rvert_{\rm t=t_1}\simeq1=\delta_i^{-1}\delta_i=\left(\frac{a_{\rm 1}}{a_i}\right)\delta_i.
\end{align}
During matter domination  $k=aH\propto a^{-1/2}$ 
so the ratio between the scale at horizon entry, $k_{i}$, and the scale at the end of matter domination, $k_1$, goes as
\begin{equation}
    \frac{k_{i}}{k_1}={\left(\frac{a_{1}}{a_i}\right)}^{\frac{1}{2}}.
\end{equation}
Therefore, the scale of the horizon needs to grow by at least $\delta_i^{-\frac{1}{2}}$ if that density fluctuation is to go on to successfully collapse. Only the most extreme fluctuations need to be given time to collapse in order to achieve the observed constraint on $\beta$ for each PBH mass. How far into the tail of the distribution we must go for each $\beta$ corresponding to a PBH mass entering the horizon at $t_i$ is given by solving for $x_{\rm tail}$
\begin{equation}
    \beta={\rm Erfc}\left(\frac{x_{\rm tail}}{\sqrt{2}}\right).
\end{equation}
For example, if $x_{\rm tail}=5$, then only 5-sigma fluctuations (i.e.~those with $\delta_i>5\times\sigma$) need to collapse in order to achieve the observed constraint on $\beta$ (appropriately scaled via equation (\ref{eqn:beta0})). We therefore require
\begin{align}\label{eqn:ineq2}
    \sigma>\frac{\left(\frac{k_{i}}{k_1}\right)^{-2}}{x_{\rm tail}}.
\end{align}
If the value of $\sigma$ from equation (\ref{eqn:powermatter}) satisfies equation (\ref{eqn:ineq2}), then there is enough time for the overdensity corresponding to that ${\sigma}$ to grow to order 1 and collapse before the end of matter domination. If a value of ${\sigma}$ does not satisfy equation (\ref{eqn:ineq2}), then the constraint on the power spectrum  must weaken to the minimum value of $\sigma$ that allows enough time for the growth of the density fluctuation to order 1 from the time of horizon crossing to the end of matter domination. This minimum value for the power spectrum of $\delta$ measured at $k=k_i$ is
\begin{equation}\label{sigma-min}
    \sigma_{\rm min}^2=\left(\frac{\left(\frac{k_{i}}{k_1}\right)^{-2}}{x_{\rm tail}}\right)^2.
\end{equation}
For each scale, we will choose the maximum of ${\sigma}$ and $\sigma_{\rm min}$. The results are plotted in figure \ref{fig:gorbonovmatter}, where we have again converted from $\mathcal{P}_\delta$ to $\mathcal{P}_\mathcal{R}$ via equation (\ref{eqn:deltaR}) with $\omega=0$, hence $\mathcal{P}_\mathcal{R}=(25/4)\mathcal{P}_\delta$.

For the shortest periods of matter domination, $t_1\sim10^{-28}-10^{-25}\,{\rm s}$, the value of ${\sigma}$ found from equation (\ref{eqn:powermatter}) is too small for the corresponding overdensity to have time to grow to order 1 before radiation domination on any scale that we have an observed value of $\beta$ for, and so the constraints are weakened. The rightmost point of each line represents the smallest PBH mass that we constrain with data from \cite{Carr:2017edp}, $10^{10}{\rm g}$. For $t_1>10^{-25}\,{\rm s}$, there begins to be sufficient time for the initial density fluctuation corresponding to ${\sigma}$ on some scales to grow to order unity before the end of matter domination. The tightest constraint is achieved for $t_1\sim10^{-19}{\rm s}$ at a scale of $k\simeq5\times10^{16}{\rm Mpc^{-1}}$, surpassing $\mathcal{P}_\mathcal{R}=2\times10^{-9}$, Planck's measurement of the amplitude of the primordial power spectrum. Constraints on the power spectrum improve for all values of $t_1$ between $10^{-28}{\rm s}$ and $10^{-6}{\rm s}$ in comparison to the constraints from PBH formation in radiation domination, with constraints for $t_1\sim10^{-5}{\rm s}$ only just overlapping with the constraint from radiation domination on some scales. The constraints due to values of $\sigma_{\rm min}$ join up with those from radiation domination as expected, since PBHs forming at the end of the early matter era will predominantly feel the effects of radiation domination if the transition occurs very soon after their formation. Note that uncertainties are introduced in the comparison between matter domination constraints and radiation domination constraints because those from radiation domination assume spherical symmetry, whereas those from matter domination do not. Additionally, for radiation domination we take $\gamma=0.2$ and $\delta_c=0.42$, but for matter domination we approximate $\gamma=1$. This explains why the radiation constraint is slightly stronger than the matter constraint on the left-hand end of each line.

We choose the latest termination of matter domination to be $t_1\sim10^{-5}{\rm s}$ for two reasons. The expression in equation (\ref{eqn:powermatter}) is only valid for $\sigma<0.05$ \cite{Harada:2016mhb}, so we cannot trust the relation between $\sigma$ and $\beta$ for $\mathcal{P}_\delta\gtrsim10^{-3}$. Secondly, the QCD phase transition occurs around $10^{-5}{\rm s}$ at an energy scale which can be probed in the laboratory, and it is generally expected that the hot big bang will be complete by this time, with the Universe dominated by radiation (although counter examples exist, see e.g.~\cite{Figueroa:2016dsc}).

We assume an instantaneous phase transition from matter domination to radiation domination, but of course the true dynamics of this transition would affect the constraints. A smooth weakening of the constraints from matter domination to match those from radiation domination is most likely. In addition, when cutting our plots at the end of matter domination, we require that the overdensities must have reached turnaround by the time $t_1$. We expect that some overdensities will have grown considerably but not quite reached order 1, however it is possible that they would still collapse to form PBHs at some point during or after the transition from matter to radiation domination - these cases have been disregarded in our constraints. Simulating the growth and subsequent collapse of the overdensity during the phase transition, as well as the dynamics of the phase transition itself, would be necessary to gain a full understanding of the effect of these cases on the constraints.

\subsection{Inhomogeneous effects}\label{section:inhom}

\begin{figure}
\centering
    \includegraphics[width=15cm]{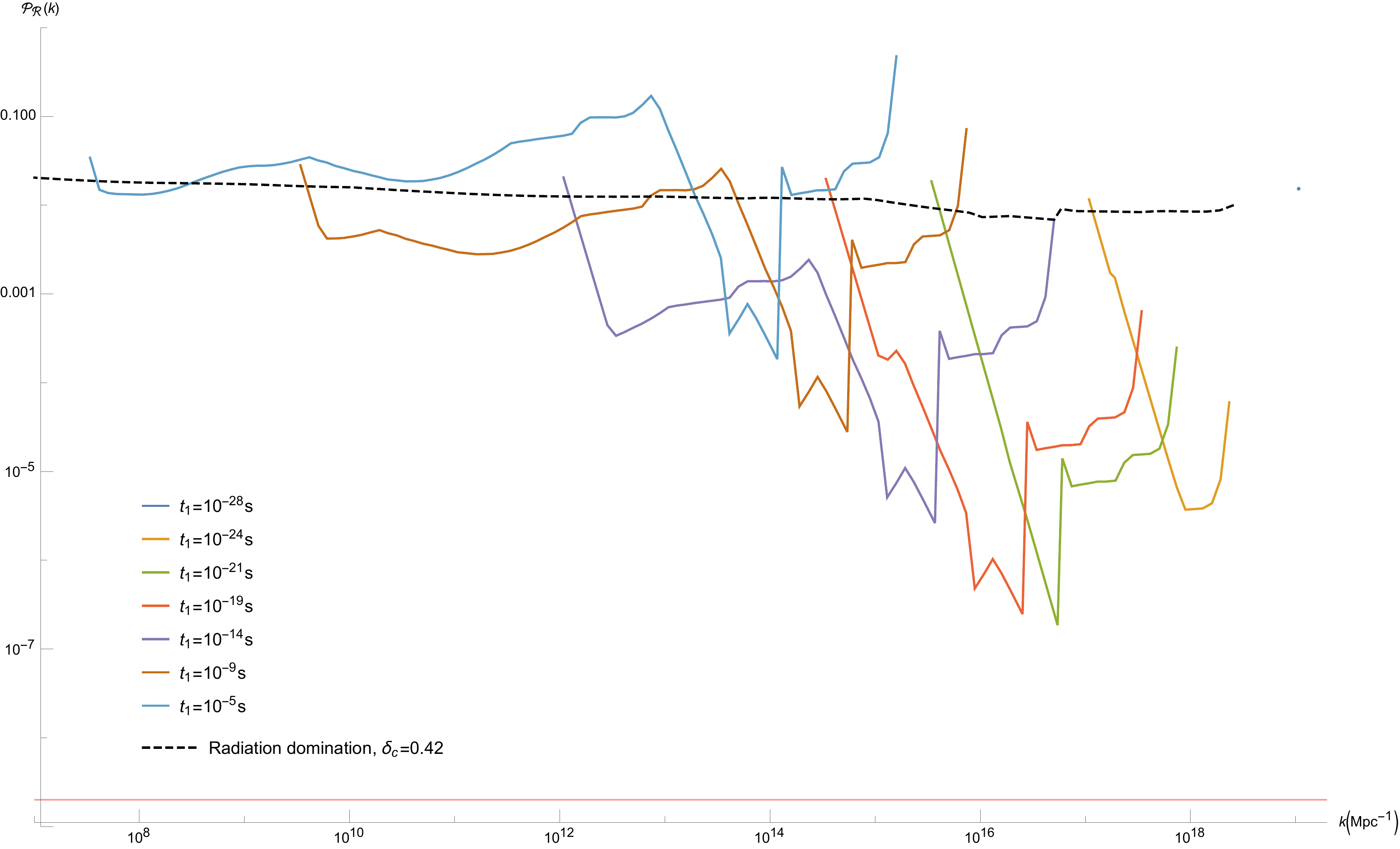}
    \caption{Inhomogeneous effects on PBH formation during matter domination translated to the power spectrum. Constraints are weakened in comparison to figure \ref{fig:gorbonovmatter} for all values of $t_1$ from $10^{-28}\,{\rm seconds
}$, represented by the single rightmost point, to $10^{-5}\,{\rm s}$, represented by the pale blue line that reaches the largest scales. The horizontal red line is drawn at $\mathcal{P}_\mathcal{R}(k)=2\times10^{-9}$. The dashed black line shows the constraints from PBHs formed in radiation domination.}
    \label{fig:inhomogeneous}
\end{figure}

The relationship  (\ref{eqn:Harada}) was derived by considering departures from spherical symmetry but neglecting potential effects of inhomogeneities in the collapsing region. If these effects were to be included, they would account for a scenario where a caustic could form in the centre of the region, and the increase in pressure could prevent a PBH from forming. The probability of this happening adds an additional factor of $\sigma^{\frac{3}{2}}$ in the relationship between $\beta$ and $\sigma$, calculated using the Lemaitre-Tolman-Bondi dust solution \cite{Polnarev1981} (see also the more recent review \cite{Khlopov:2008qy}). With the most conservative reasoning, this effect is considered to be independent of the probability that the region is spherical enough to collapse into a PBH rather than a pancake or cigar shape, which accounts for the factor of $\sigma^5$ that we have been using \cite{Doroshkevich,Polnarev1981}. Multiplying these gives the minimum probability of PBH formation to be
\begin{equation}\label{eqn:inhom}
    \beta_0\simeq2\times10^{-2}\sigma^{\frac{13}{2}}.
\end{equation}
Using this relationship originally from \cite{Polnarev1981,Doroshkevich} instead of (\ref{eqn:Harada}), we can plot the power spectrum vs.~scale again, shown in figure \ref{fig:inhomogeneous}. The strongest constraint on $\beta$ now produces a constraint on the power spectrum of order $\mathcal{P}_{\mathcal{R}}\simeq10^{-7}$, two orders of magnitude weaker than if the inhomogeneous effects are neglected. Additionally, values of $\sigma$ become larger than 0.05 which were considered not valid for the equation (\ref{eqn:Harada}) in \cite{Harada:2016mhb}, however \cite{Polnarev1981} does not cite this as a limiting factor of the equation that accounts for inhomogeneous effects given in equation (\ref{eqn:inhom}). We caution against concluding that the constraints in a matter dominated era may be weaker than those in radiation domination (which would be very surprising since pressure can only act against gravitational collapse); the radiation era constraints are derived assuming spherical symmetry which maximises the probability of PBHs forming. 

Arguments for neglecting the effect of inhomogeneities include that they are very dependent on the matter model, and since we have not specified what has caused the early matter dominated phase, these effects are quite uncertain. Additionally, it was argued in \cite{Harada:2016mhb} that pressure arising in the central region could just slow down the collapse as opposed to halting it completely. For other caveats relating to the use of this formula see \cite{Gorbunov:2017fhq}.

\section{No PBHs and an early-matter phase}\label{section:mossmatter}

When the two scenarios explored so far in sections \ref{Ian Moss} and \ref{Matter} are combined such that there is presumed to be a period of early matter domination during which no PBHs of masses up to $10^{15}\,{\rm g}$ form, because they would have otherwise seeded rapid vacuum decay, constraints on the power spectrum tighten by many orders of magnitude. Using the same argument from equation (\ref{betaconstr}) that the Universe can be split up into regions of the scale of a potential PBH, we can reformulate our expression relating the power spectrum to the PBH mass fraction so that it reads
\begin{equation}\label{eqn:Nlmatter}
    \mathcal{P}_\delta\simeq\left(\frac{1}{0.056N_l}\right)^{\frac{2}{5}},
\end{equation}
using the relationship from \cite{Harada:2016mhb} that disregards inhomogeneous effects. If we were instead to account for inhomogeneous effects and use the more conservative expression (\ref{eqn:inhom}) the constraints would weaken similarly to the case in section \ref{section:inhom}.
\vspace{5mm}
\begin{figure}
\centering
\includegraphics[width=15cm]{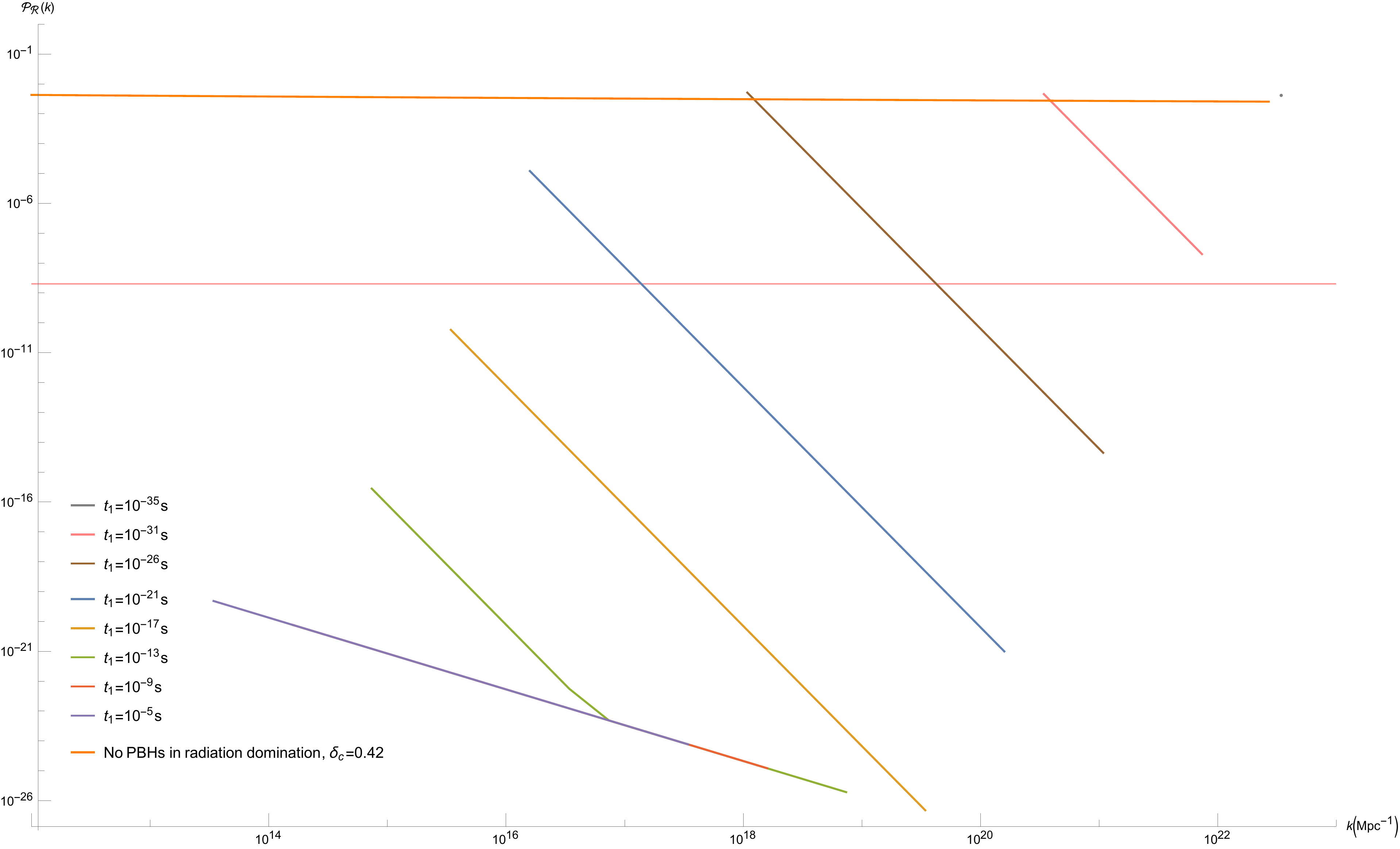}\captionof{figure}{The upper bound on the power spectrum assuming that no PBHs have ever decayed.
The horizontal red line is the power spectrum amplitude measured on CMB scales. For the lines in matter domination, the left hand side of the line corresponds to the scale when the horizon mass is $10^{15}{\rm g}$ (which corresponds to the heaviest PBH which would have decayed by today) and the right hand side corresponds to the scale when the horizon mass is $10^3{\rm g}$ (which is the lightest PBH mass that could form after inflation assuming that approximately 60 efolds occur between today's horizon scale exiting the Hubble sphere during inflation and the end of inflation). The nearly horizontal orange line is the constraint from no PBHs forming in radiation domination for $\delta_c=0.42$. Notice that a much smaller range of scales is being plotted here compared to all previous plots showing the constraints on the power spectrum.}\label{fig:mossmatter}
\end{figure}

Figure \ref{fig:mossmatter} demonstrates how the constraints on the primordial power spectrum tighten by many orders of magnitude when the two scenarios are combined. The right hand side of each line is the smallest scale plotted for each value of $t_1$, corresponding to the scale when the horizon mass is $10^3{\rm g}$. This is the lightest PBH mass that could form after inflation assuming that approximately 60 efolds occur between today's horizon scale exiting the Hubble sphere during inflation and the end of inflation. Horizon entry time for a PBH mass of $10^3{\rm g}$ is $t_i\simeq3\times10^{-36}{\rm s}$, so the earliest value of $t_1$ we use is $t_1=10^{-35}{\rm s}$. Note that allowing $t_1$ to vary either causes the energy scale of inflation to vary if the number of e-foldings between today's horizon scale exiting the Hubble sphere during inflation and the end of inflation is kept fixed, or varies the number of e-foldings if the energy scale of inflation is kept fixed. Since there is no obvious benefit to one or the other choice, the number of e-foldings is kept fixed in this work. Only constraints due to PBHs of masses small enough that they would have decayed by today are plotted in figure \ref{fig:mossmatter}, so the left hand side of each line corresponds to the scale when the horizon mass is $10^{15}{\rm g}$. 

For $t_1<10^{-16}{\rm s}$, initial fluctuations corresponding to values of $\sigma$ from equation (\ref{eqn:Nlmatter}) are too small on all scales that enter the horizon before the end of the matter phase to grow to order 1 before radiation domination, so $\sigma_{\rm min}$ is always chosen. For $t_1$ greater than this, constraints from equation (\ref{eqn:Nlmatter}) start to become valid. The best constraint is achieved at a scale of $k\simeq5\times10^{19}{\rm Mpc^{-1}}$ with $t_1\sim10^{-17}{\rm s}$, and reaches $\mathcal{P}_\mathcal{R}\simeq5\times10^{-27}$. Models which generate a nearly scale-invariant spectrum are ruled out for values of $t_1$ between $10^{-5}{\rm s}$ and $10^{-30}{\rm s}$ as constraints on the power spectrum tighten further than an amplitude of $\mathcal{P}_\mathcal{R}=2\times10^{-9}$ (as measured by Planck) on some portion of scales depending on the value of $t_1$. If constraints were required for the situation where no PBHs of any mass are formed, the lines plotted in figure \ref{fig:mossmatter} can be extrapolated from the left-most point of each line to values of around $\mathcal{P}_\mathcal{R}\sim10^{-2}$, depending on the value of $t_1$. PBHs forming right at the end of a matter-dominated phase will essentially only be affected by the radiation dominated phase that begins soon after, so it is unsurprising that the constraints become similar to those from no PBHs forming in a radiation dominated background (shown by the nearly horizontal orange line). The constraints from radiation domination are slightly stronger because they assume spherical collapse, whereas the constraints from matter domination do not, and similarly we choose $\delta_c=0.42$ for radiation domination but only that the perturbation must grow to order 1 for matter domination. 

Such extreme constraints on the power spectrum suggest that an early-matter phase is incompatible with no light PBHs forming during this phase. If the power spectrum is quasi-scale invariant over all scales, then only a sufficiently brief matter dominated period of less than about 7 efoldings is allowed, as the density perturbations would not have time to collapse into a PBH before radiation domination. A complementary approach was presented recently in \cite{Gorbunov:2017fhq}, showing that it is possible that the probability of PBHs nucleating rapid vacuum decay has been overestimated, or otherwise constraints on the energy level at the end of inflation must be enforced to avoid light PBHs forming if an early matter dominated phase occurred. They show that if the energy density at the end of inflation is less than $(2\times10^9{\rm GeV})^4$, then the first and hence lightest PBHs produced after inflation will be sufficiently massive to not have decayed. The constraints found on the energy scale of inflation and reheating assume a flat power spectrum, and hence put a constraint on the duration of the early-matter dominated phase. In this work, the duration of the early-matter phase has been allowed to vary and hence the constraints are cast in terms of upper bounds on the power spectrum instead. % 
Even more recently, \cite{Canko:2017ebb} show that a nonminimal, but renormalizable coupling between the Standard Model Higgs field and gravity can have a large effect on the decay rate of the vacuum, possibly allowing light PBHs to form without nucleating decay to the true vacuum.

\section{Conclusion}

PBHs constrain the primordial power spectrum over an extremely broad range of scale, covering over 20 orders of magnitude. However, the constraints are much weaker than the observed amplitude on CMB scales, due to the high pressure forces during radiation domination, which mean only very large amplitude perturbations can collapse. In this paper, we have considered two possible scenarios to tighten the constraints. Firstly we consider the ultimate observational constraint that no PBHs formed during a standard radiation era and secondly the softest possible equation of state - an early matter dominated era which makes PBH formation much easier. Finally we combine the two scenarios. 

Although improving the observational constraints is very important for potentially ruling out PBHs as a dark matter candidate (see for example \cite{Kovetz:2017rvv,Garcia-Bellido:2017fdg,Garcia-Bellido:2017mdw,Kuhnel:2017pwq,Georg:2017mqk,Niikura:2017zjd,Inomata:2017okj}), we show in Fig.~\ref{fig:ianmoss} that the consequent improvement in the constraints on the primordial power spectrum is modest, being less than an order of magnitude. The only exception is on very small scales corresponding to PBHs which form with such small masses that they decay before big bang nucleosynthesis (corresponding to $k\gtrsim 10^{18}/ \rm{Mpc}$), on these small scales there are no standard observational constraints. However, the argument that the evaporation of any black hole could destroy the stability of the Universe \cite{Burda:2015isa,Burda:2016mou} suggests that arbitrarily small scales (right down to the horizon scale at the end of inflation) can be constrained. The power spectrum constraints are so insensitive to changes in $\beta$ of even 50 orders of magnitude, that a better understanding of the formation of PBHs (e.g.~simulations of non-spherical initial conditions or reaching a better understanding of the expected initial density profile of PBHs from inflation) will have a larger effect than improving observational constraints. Non-Gaussianity of the primordial density perturbation has already been shown to effect the constraints by up to two orders of magnitude. 

However, we have shown that in the matter domination case, due to the enhanced probability of PBH formation, the constraints on the power spectrum can be improved by many orders of magnitude and they become much more sensitive to the observational constraints on $\beta$. The constraints also depend on the length of the early matter dominated phase, as shown in Fig.~\ref{fig:gorbonovmatter}. For an early matter dominated phase lasting $10^{-19}\,{\rm s}$, the constraint on the primordial power spectrum surpasses Planck's measurement of $\mathcal{P}_\mathcal{R}=2\times10^{-9}$ at a scale of $k\simeq5\times10^{16}\,{\rm Mpc^{-1}}$.
We further include full constraints for scales entering the horizon right up until the end of matter domination, which consistently match with those from radiation domination to within the uncertainties between the two calculations. If the constraints were to tighten further than this on a wider range of scales, models of inflation that generate a quasi scale-invariant spectrum could   be ruled out when combined with an early-matter dominated phase.

In combining these two scenarios, that no PBHs that would have evaporated by today can ever have formed, and that there was a phase of early-matter domination post inflation, we show that incredibly tight constraints on the power spectrum can be achieved for some lengths of a matter dominated phase. If both scenarios can be shown to be realised in nature, then this provides an excellent direction for highlighting the most probable models of inflation, the power spectrum of which would need to decrease by many orders of magnitude on small scales. This may suggest that one or both these scenarios are not realised, or that the energy scale at the end of inflation must be so low such that sufficiently light PBHs that would have decayed by today could not have formed \cite{Gorbunov:2017fhq}. 
\newline\newline
\indent {\bf Note added:}  Whilst our paper was being written, \cite{Carr:2017edp} produced a similar plot to Fig.~\ref{fig:gorbonovmatter}, showing the constraints on $\mathcal{P}_\mathcal{R}$ for a radiation and matter dominated era. Our constraints are an order of magnitude weaker than theirs, which we believe is primarily due to a factor of 9 difference in the conversion between $\delta$ to $\mathcal{R}$. We use eq.~(\ref{eqn:deltaR}) evaluated at horizon entry, while they relate $\delta$ and $\mathcal{R}$ on super-horizon scales. They also do not consider constraints due to $\sigma_{\rm min}$ (see eq.~({\ref{sigma-min})), which was not important for their purposes. We note that our constraint on scales which enter the horizon during radiation domination of $\mathcal{P}_\mathcal{R}\lesssim10^{-2}$ agrees with \cite{Josan:2009qn}.

Two papers relating the vacuum stability argument that no PBHs may have evaporated during a matter dominated phase also recently appeared \cite{Gorbunov:2017fhq,Canko:2017ebb}, we comment on these papers in Sec.~\ref{section:inhom}.

\section*{Acknowledgements}
The authors thank Tommi Tenkanen for helpful discussions. PC acknowledges support from the UK Science and Technology Facilities Council via Research Training Grant ST/N504452/1, CB is supported by a Royal Society University Research Fellowship.

\bibliography{extreme}
\end{document}